\documentclass[twocolumn,showpacs,preprintnumbers,amsmath,amssymb,prl]{revtex4}

\usepackage{graphicx}
\usepackage{dcolumn}
\usepackage{bm}

\newcommand{\etal}{{\em et al.}}

\begin{document}

\title{First-principles study of the Young's modulus of Si $<$001$>$ nanowires}

\author{Byeongchan Lee}
\author{Robert E. Rudd}
 \email{robert.rudd@llnl.gov}
\affiliation{Lawrence Livermore National Laboratory, University of California, L-415, Livermore, California 94551}

\date{\today}

\begin{abstract}
We report the results of first-principles density functional theory
calculations of the Young's
modulus and other mechanical properties of hydrogen-passivated
Si $<$001$>$ nanowires.  The nanowires are taken to have predominantly
\{100\} surfaces, with small \{110\} facets.
The Young's modulus, the equilibrium length and the residual
stress of a series of prismatic wires are found to
have a size dependence that scales like the surface area to volume
ratio for all but the smallest wires.
We analyze the physical origin of the size dependence, and compare
the results to two existing models.
\end{abstract}

\maketitle


Nanoscale mechanical devices have been proposed
for applications ranging from nano-electro-mechanical
systems (NEMS) such as high frequency
oscillators and filters\cite{SiResonators1}
to
nanoscale probes\cite{nanotubeProbe}
to
nanofluidic valves\cite{WAG}
to
q-bits for quantum computation\cite{ClelandGeller}.
The process of design and
fabrication of these devices is extremely challenging, complicated
in part by uncertainties about how even ideal devices should behave.
The mechanical response of structures at the nanoscale is known
to be different than that of their macroscopic analogs and surface
effects in these high surface-to-volume devices are important\cite{Cahn},
but a predictive theory of nanomechanics remains an open problem.

%

Much of what is known about mechanics of nanodevices has been
learned from atomistic calculations based on empirical potentials.
The first such calculations were done for
single-crystal alpha quartz beams, finding that the Young's modulus
decreased with decreasing size\cite{BMVK,quartzResonators}.
These and calculations of the Young's modulus for various
other materials have found a size-dependent modulus
with an additive correction to the bulk value that
scales like the surface area to volume ratio\cite{Miller,RuddIJMSE}.
A few studies
claim an additional contribution that scales like the edge
to volume ratio (cf.\ \onlinecite{BMVK}), and such a contribution, with
a factor of the logarithm of the separation of
the edges, has been discussed for
epitaxial quantum dots\cite{Shchukin,RuddQDot}.
An intuitive way of understanding these effects
is that there is a layer of material at the surface (and edges)
whose mechanical properties differ from those of the bulk
including different elastic moduli and eigenstrains.
This layer could be chemically
distinct from the bulk, such as an oxide layer or a
hydrogen-passivated surface, but the effect may be entirely due to
the structural difference at the surface, such as a bare
reconstructed surface.
The formalism of nanoscale mechanics
based on the surface energy and its first two strain derivatives
(the surface stress and modulus) has been
developed\cite{Miller,Kukta}.
Recently it has been proposed that the size dependence of the Young's
modulus can be due to the anharmonicity (non-linearity) of the
bulk elastic moduli together with the strain resulting from the
surface stress\cite{bulkAnharmonicity}.

To date experimental data on the size dependence of nanostructure
mechanics are very limited.  Atomic force microscopy (AFM) measurements
of the Young's modulus ($E$)\cite{LieberMechanics}
of cast metallic nanowires show a strong size
dependence\cite{Cuenot}.
Recent experiments have also found a strong size dependence for
$E$ of ZnO nanowires\cite{ZnOnanowire},
and other mechanical properties
of ZnO and GaN nanowires\cite{Nixnanowire}.
Measurements of $E$ for silica
nanobeams have demonstrated that the way in which the beam is clamped
(i.e.\ the boundary conditions) affects the apparent value\cite{Ruoff2}.
A study using a different AFM technique reported a value of $E$
of $18\pm2$ GPa for a $<10$ nm Si [100] nanowire\cite{Kizuka};
for 100-200 nm Si$\{$111$\}$ wires, $E$ has been found to be consistent
with the bulk value\cite{PYangSiNW}.
Experimental challenges measuring the intrinsic nanoscale Young's modulus
make this a topic of continued activity, leveraging earlier work on the
mechanics of nanotubes\cite{nanotubeMech}.

In the absence of definitive experimental data, first-principles quantum
mechanical calculations can provide robust predictions of nanowire
mechanical properties, but
few results have been reported.  One quantum study based on an empirical
tight-binding technique has been published\cite{Arias}.
The electronic and optical properties
of nanowires have been studied using first-principles techniques,
leading to interesting predictions about size-dependent phenomena
as evidenced by an increase in band gap due to
quantum confinement\cite{confinement1,confinement2} and
a switch from an indirect, to a direct,
band gap\cite{electronicSiNW1,electronicSiNW2}.
We are not aware of any {\em ab initio} calculations of nanowire
moduli.

Here we present first-principles calculations of the
mechanical properties of silicon nanowires,
studying the Young's modulus
due to its direct relevance in the function of nanoscale devices
such as flexural-mode mechanical resonators\cite{SiResonators1}
and as an archetype for a variety of mechanical properties.
We address several important open questions in nanomechanics.
Is the modulus size dependent? Does it soften or stiffen at the
nanoscale?  What physics cause the effect?
We focus on prismatic Si $<$001$>$ nanowires with a combination of
$\{$100$\}$ and $\{$110$\}$ H-passivated surfaces, and
single crystal cores as in
experiment\cite{SiResonators1,SiResonators2}. We have
chosen the [001] orientation for the longitudinal axis because
of its relevance to the NEMS devices\cite{SiResonators1};
Si nanowires grown rather
than etched typically have different orientations\cite{Lieber}.
Hydrogen passivation results from rinsing the oxidized
Si surfaces with HF, and it provides a standard system suitable for
a systematic study of size dependence in nanomechanics.
With other surface conditions the band gap can vary greatly, and nanowires
can go from semiconducting to metallic\cite{metallicSiNW};
but the H-passivated wires remain
semiconducting\cite{SiNWSurface} and the surfaces do
not change the nature of
Si-Si chemical bonding from its covalent character.


First-principles density functional theory (DFT) has been employed:
specifically, the Vienna Ab-initio Simulation Package
using the projector augmented-wave method\cite{PAW1,PAW2}
within the generalized gradient approximation (GGA)\cite{GGA}.
The energy cutoff for the plane wave expansion is 29.34~Ry and higher,
and 6 points in the one-dimensional irreducible Brillouin zone are
used for k-point sampling. Each supercell is periodic,
is one Si cubic unit cell long
along the wire
and has more than 10~{\AA} vacuum space in the transverse directions.


To calculate $E$, we define the cross-sectional area to be the area
bounded by the centers of the outermost (H) atoms.
This choice is motivated by the fact that the volume excluded
by the wire from access by outside atoms is determined from
the forces arising from electron interactions.
Most of the electron density is enclosed by the
boundary formed by H atoms and the electron density from
Si atoms essentially vanishes beyond this point. The
positions of the nuclei are well defined and not subjective.
Other definitions of the bounding surface exist, for example
the mid-plane between two identical H-passivated surfaces at
their minimum energy separation\cite{SiSlabs}.

\begin{figure}
\includegraphics[width=0.45\textwidth]{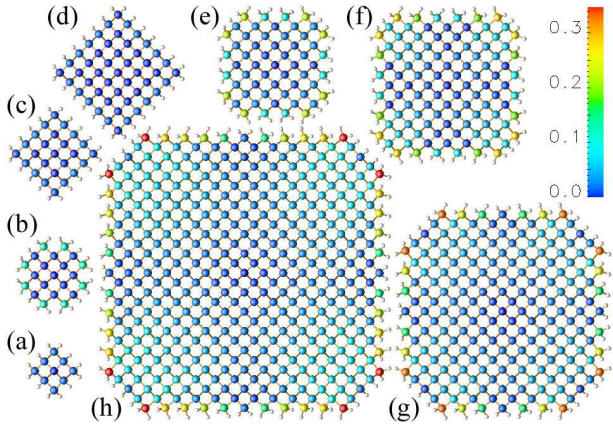}
\caption{\label{fig:wires} (Color online)
Cross-sections of fully relaxed hydrogen-passivated wires,
with each Si atoms colored as shown in the legend corresponding
to its transverse relaxation in \AA.
The widths of wires are (a) 0.61 (b) 0.92, (c) 1.00, (d) 1.39,
 (e) 1.49, (f) 2.05, (g) 2.80, and (h) 3.92 nm respectively.
The width is defined as square-root of the cross-sectional area.}
\end{figure}

\begin{figure}
\includegraphics[width=0.45\textwidth]{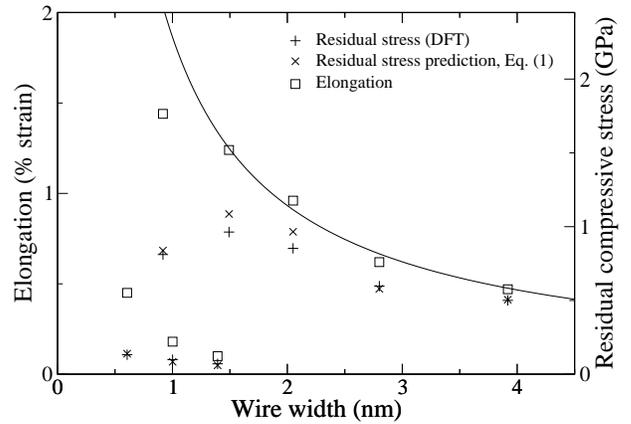}
\caption{\label{fig:elongation}Silicon nanowire axial stress and
equilibrium elongation strain calculated in DFT as a function of wire size.
The solid curve is a fit to $C/w$ of the elongation strain
to 4 data points from 1.49~nm and bigger wires, with $C$=1.9\%-nm.
The predictions of Eq.\ \ref{eq:resStress} are also plotted using
the stresses from DFT calculations of hydrogenated 14-layer $\{100\}$ and 15-layer $\{110\}$ slabs.
The asterisk-like symbols are from overlapping $+$ and $\times$ symbols.}
\end{figure}

The cross-sectional shape of the Si [001] wire is a truncated square
with four $\{$100$\}$ facets and four $\{$110$\}$ facets. Some
wires studied have no $\{$100$\}$ facets; for those that do, the ratio
of the facet areas is taken to be roughly in accordance with the
Wulff shape for a bare wire with (110)-(1x1) and (100)-p(2x2) surface
reconstructions; i.e.\ the ratio of $\{$100$\}$ to $\{$110$\}$ area is 3.5:1.
For each of the nanowire geometries shown in Fig.\ \ref{fig:wires},
the Si atoms were initially
positioned at their bulk lattice sites and hydrogens were added
to terminate the bonds at the surfaces, and this configuration
was relaxed.
%
The system was relaxed
to its zero-temperature minimum energy with the
length of the periodic supercell held fixed at the bulk value in
the longitudinal direction.  The axial stress in this configuration,
$\sigma _{zz} (L_0) =
  \left. V^{-1} \partial U/\partial \epsilon _{zz}\right| _0$
where $U$ is the DFT total energy,
is indicative of
the residual stress in a doubly clamped beam etched from a
single-crystal substrate.
It is plotted in Fig.\ \ref{fig:elongation}.

\begin{figure}
\includegraphics[width=0.45\textwidth]{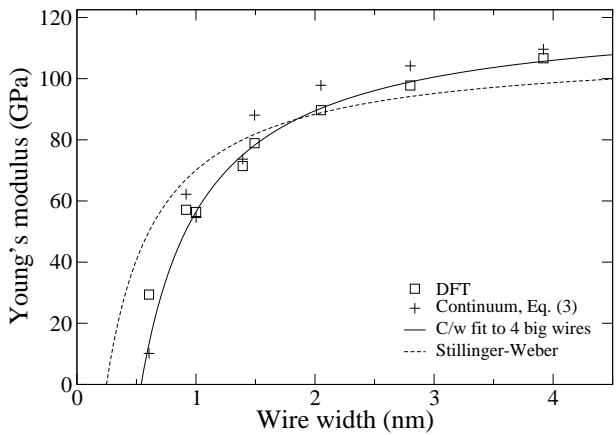}
\caption{\label{fig:modulus} Silicon nanowire Young's modulus
calculated in DFT as a function of wire size.
For comparison values of continuum formula (\ref{continuumE})
are also plotted,
using the $\{100\}$ and $\{110\}$ surface elastic constants
obtained in DFT from hydrogenated 14-layer $\{100\}$ and 15-layer $\{110\}$ slabs respectively.
The solid curve, $E=E_{bulk}^{{\mathrm{DFT}}} - C/w$ with
$C$=66.11 GPa/nm, is the best fit to a pure surface area to volume
size dependence.}
\end{figure}

Next the relaxed total energy was calculated for each wire in a series
of longitudinal strains, at increments of roughly 0.5\%.  These total
energy values were fit to a polynomial.  The minimum of the polynomial gives
the equilibrium length, and the value of the curvature at the
minimum gives the Young's modulus,
$E=\left. V^{-1} \partial ^2 U/\partial \epsilon _{zz}^2\right| _{\epsilon_{zz-min}}$.
The equilibrium elongation and modulus are plotted in
Figs.\ \ref{fig:elongation} and \ref{fig:modulus}, respectively.

The calculated Young's modulus of the 1.49~nm wire is
tabulated in Table~\ref{tab:fitting}. The table gives an indication of the sensitivity
to the order of the polynomial fit.
For the given order, a
higher cutoff energy offers little improvement.
We find that the second order fit with 29.34 Ry energy cutoff
is reasonably good, differing by less than 2\% compared with
all the higher-order combinations tested. The second-order fit also
permits direct comparison with the results from bigger wires where
the number of data points and the energy cut off are limited by
the computational cost of systems up to 405 Si and 100 H atoms.

\begin{table}
\caption{\label{tab:fitting}The calculated Young's modulus in GPa of the
1.49~nm nanowire
as a function of the
plane wave cutoff energy and the order of the fit.
The same 10 data points were fit for each polynomial order.}
\begin{ruledtabular}
\begin{tabular}{cccccc}
Cutoff energy & \multicolumn{5}{c}{Order of polynomial fitting}\\
(Ry) & 2nd & 3rd & 4th & 5th & 6th\\
\hline
29.34 & 78.90 & 78.88 & 79.90 & 79.94 & 78.61\\
44.10 & 79.31 & 79.28 & 80.39 & 80.33 & 78.95\\
51.45 & 79.40 & 79.37 & 80.35 & 80.31 & 79.01
\end{tabular}
\end{ruledtabular}
\end{table}


These calculations allow us to analyze the physical origin of the size
dependence.
Size dependences of the residual stress and the elongation
evident in Fig.\ \ref{fig:elongation}
are driven by the same physics: compressive surface stress.
The residual axial stress of the Si beam may be decomposed into
core, H adatom and Si surface parts:
core contributions from the Si atoms,
extrinsic contributions from hydrogen (H-H) interactions
and intrinsic surface
contributions from the change to the Si bonds near
the surface compared to the Si bulk (Si-H and modified bond order Si-Si).
Since DFT only provides a total energy, this decomposition is
somewhat ambiguous.  We estimate the H-H interactions as equal to
those of neighboring hydrogens in two silane molecules in the
orientation and separation of the H-passivated surface,
and the core contribution to be the axial stress in bulk Si
uniformly strained to match the nanowire;
the intrinsic contribution is the remainder.
The extrinsic contribution is most important, as we now show.
The intrinsic surface stress is small, as expected since the dangling
Si bonds are well terminated with H and the Si-Si bond order is not
significantly different than in the bulk.  The small magnitude of the
intrinsic stress is best seen in the case of the 1.39~nm
wire for which the elongation is less than 0.1\% compared to $\sim$1.5\%
of the 1.49~nm wire.  The absence of $\{$100$\}$ facets on this wire
leads to a small extrinsic stress since the H-H separation on the
$\{$110$\}$ facets is relatively large.
The vacant Si sites above the facets are filled by one
and two H atoms on $\{$110$\}$ and $\{$100$\}$, respectively,
and the double occupancy, albeit with
$\sim$2~{\AA} H-H separation due to the shorter Si-H bond, leads to more
repulsion for $\{$100$\}$\cite{BCLeeInPrep}.

The extrinsic surface stress due to the H-H repulsion
on the $\{$100$\}$
facets quantitatively accounts for both the compressive residual stress $\sigma _{zz} (L_0)$
and the elongated equilibrium length $L_{eq}$ of the nanowires.
They are
related to leading order through the linear elasticity:
\begin{eqnarray}
\sigma _{zz} (L_0) & = & \sigma _{zz} (core) +
\frac{1}{A} \sum _i \tau ^{(i)}_{zz} w_i
\label{eq:resStress}
\\
\left( L_0 - L_{eq} \right) / L_{eq} & \sim & \sigma _{zz} (L_0) / E
\label{eq:eqLen}
\end{eqnarray}
where $A$ is the cross-sectional area, $w_i$ is the width,
$\tau ^{(i)}_{zz}$ is the longitudinal surface
stress of facet $i$, and $L_0$ is the bulk length of the beam.
$E$ is the Young's modulus of the beam.
For constant surface stress, the second term in Eq.\ (\ref{eq:resStress})
is proportional to the surface area to volume ratio; the core
stress is too, since the surface stress causes a transverse
expansion of the wire that induces a tensile core stress.
We now use much smaller periodic slabs to quantify these
contributions and compare with the nanowire results.
Using H-passivated slabs we calculate in DFT
the surface stress of the ground-state canted (100)
surface to be -55.0 meV/\AA$^2$, and that of
the (110) surface
to be -1.3 meV/\AA$^2$\cite{surfaceEfootnote}.
The negative stress indicates compression.
The core stress may be estimated through a generalized Young-Laplace
law to be
$\sigma _{zz} (core) \approx - 8 \nu \tau ^{\{100\}}_{zz}/ \pi w$,
where $\nu = C_{12}/(C_{11}+C_{12})$ is the Poisson ratio.
The details of these calculations will be given
elsewhere\cite{BCLeeInPrep}.
Using these values in Eq.\ (\ref{eq:resStress})
gives predictions in very good agreement with the full nanowire
calculations as shown in Fig.\ \ref{fig:elongation}.
The scatter
for 1.49 and 2.05~nm wires may be accounted for by small edge effects.
The 0.61, 1.00 and 1.39~nm wires have no $\{$100$\}$ facets and
almost no elongation as described above.
In the case of the second smallest (0.92~nm) wire all of the $\{$100$\}$
atoms undergo substantial relaxation, as shown in Fig.\ \ref{fig:wires},
lowering the magnitude of the surface stress and the elongation.
This high level of agreement gives us confidence that
we understand the physics of the size dependence of the
residual stress.

What about the Young's modulus?
As shown in Fig.\ \ref{fig:modulus}
it becomes softer monotonically as the size is
decreased. It drops from the bulk value
($E_{bulk}^{{\mathrm{DFT}}}$ = 122.53 GPa)
in proportion to the surface area to volume ratio.
It does not exhibit the strong dependence on the ratio of $\{$100$\}$
to $\{$110$\}$ area seen in the equilibrium length.
As with the residual stress, the Young's modulus may be
decomposed into intrinsic, core and extrinsic contributions.
From continuum mechanics neglecting edge and non-local effects,
the modulus can be expressed, slightly generalizing Ref.\ \onlinecite{Miller},
as
\begin{equation}
E = E(core) + \frac{1}{A} \sum _i S^{(i)} w_i
\label{continuumE}
\end{equation}
where $S^{(i)}$ is the surface elastic constant, a strain-derivative of
the surface stress including both extrinsic and intrinsic parts.
The insensitivity to the facet ratio suggests several conclusions:
The extrinsic contribution to the modulus
(which is strongly facet dependent)
is small;
the core anharmonicity is irrelevant since the modulus
is not correlated with the equilibrium elongation; and the intrinsic
surface contribution dominates and its $\{$100$\}$ value may
be nearly sufficient to determine $E$.
To study the core stress further,
we calculated that the Young's modulus of the bulk crystal increases
by only 1.6\% when strained
$\sim$1.5\% to match the most strained (0.92~nm) wire.
This change is negligible compared to the observed
softening (contrary to claims that the bulk anharmonicity
is dominant\cite{bulkAnharmonicity}).
The extrinsic effect is also small, but not negligible.  Based
on silane interaction forces for the canted $\{100\}$ surface
geometry we have estimated that the extrinsic
contribution is $\sim$8 GPa for the 1.49~nm wire\cite{BCLeeInPrep},
roughly equal to E(1.49nm)-E(1.39nm), i.e.\ the difference
in the moduli with and without $\{100\}$ facets.
We have also calculated
the size dependence of the modulus using Eq.\ (\ref{continuumE})
based on the surface elastic constant $S^{\{100\}}$ from a separate slab
calculation\cite{BCLeeInPrep}.
The results, shown in Fig.\ \ref{fig:modulus}, are in good
agreement with the full first-principles calculation, and
adding the core contribution slightly improves the agreement.
Also plotted in the figure is the best fit curve of
Ref.\ \onlinecite{Miller} from Stillinger-Weber (SW) empirical
molecular statics calculations. The SW bulk Young's
modulus is 13\% lower and the coefficient $C$ of the $1/w$ term
is 29\% lower than the DFT values. The errors compensate
for each other leading to reasonable agreement for the
nanoscale wires, which is unexpected since the SW
potential does not have the relevant nano-physics in its
functional form or its fitting database.

In conclusion we have found that calculation of several mechanical properties
of silicon wires reveals a size dependence at the nanoscale,
allowing analysis of the magnitude of surface and edge
effects in the nanowire Young's modulus from first principles
for the first time.
In each case the size dependence scales roughly as the
surface area to volume ratio,
but for different reasons.
For the equilibrium length and residual
stress it is due to the extrinsic surface stress from interactions
in the H passivation layer; for the Young's modulus it arises from
the intrinsic contribution to the surface elastic constant.
Surface parameters from slab calculations capture most, but not all,
of the physics.
The size effect is not strong for the H-terminated surfaces studied here:
the Young's modulus is softened by about 50\% for a 1~nm diameter wire.
It may be possible to measure this effect directly using either
AFM deflection or resonant frequency measurements
in a double clamped configuration.
Another interesting possibility is that the effect could be substantially
stronger in silicon nanowires with different surfaces, such as bare
or oxide surfaces, making measurement easier.  For those systems, the
balance of core, intrinsic and extrinsic contributions could be different,
and indeed, new functional forms may be needed for the smallest wires.


We are grateful to A.\ J.\ Williamson for helpful comments.
This work was performed under the auspices of the US Dept.\ of Energy
by the Univ.\ of California, Lawrence Livermore National Laboratory,
under Contract No. W-7405-Eng-48.


\begin{thebibliography}{99}
\bibitem{surfaceEfootnote}
For comparison
we calculate the $\{$100$\}$ and $\{$110$\}$ surface energies
to be 19.6 meV/\AA$^2$ and 8.2 meV/\AA$^2$, respectively.
\end{thebibliography}
\end{document}